\documentclass[aps,prl,twocolumn,superscriptaddress,showpacs,showkeys]{revtex4-1}
\usepackage{graphicx}
\usepackage[abs]{overpic}
\usepackage{upgreek}
\usepackage{amssymb}
\usepackage{pdfpages}
\begin{document}
\newcommand{\nv}{NV$^-$}
\newcommand{\ucb}{Department of Physics, University of California, Berkeley, CA 94720-7300, USA}
\newcommand{\krakow}{Institute of Physics, Jagiellonian University, Reymonta 4, 30-059 Krak\'{o}w, Poland}
\title{Microwave saturation spectroscopy of nitrogen-vacancy ensembles in diamond}
\author{P. Kehayias}
\email[]{pkehayias@berkeley.edu}
\affiliation{\ucb}
\author{M. Mr\'{o}zek}
\affiliation{\krakow}
\author{V.M. Acosta}
\affiliation{\ucb}
\affiliation{Google [x], 1600 Amphitheatre Parkway, Mountain View, CA, USA}
\author{A. Jarmola}
\affiliation{\ucb}
\author{D.S. Rudnicki}
\affiliation{\krakow}
\author{R. Folman}
\affiliation{Department of Physics, Ben-Gurion University of the Negev, Be’er Sheva, Israel}
\author{W. Gawlik}
\affiliation{\krakow}
\author{D. Budker}
\email[]{budker@berkeley.edu}
\affiliation{\ucb}
\affiliation{Helmholtz Institute, JGU, Mainz, Germany}
\date{\today}
\begin{abstract}
Negatively-charged nitrogen-vacancy (NV$^-$) centers in diamond have generated much recent interest for their use in sensing. The sensitivity improves when the NV ground-state microwave transitions are narrow, but these transitions suffer from inhomogeneous broadening, especially in high-density NV ensembles. To better understand and remove the sources of broadening, we demonstrate room-temperature spectral ``hole burning" of the NV ground-state transitions. We find that hole burning removes the broadening caused by magnetic fields from $^{13}$C nuclei and demonstrate that it can be used for magnetic-field-insensitive thermometry.
\end{abstract}

\pacs{78.47.nd, 76.70.Hb}
\keywords{Nitrogen-vacancy centers, saturation spectroscopy}
\maketitle
\begin{figure}[t]
\begin{center}
\begin{overpic}[width=0.48\textwidth]{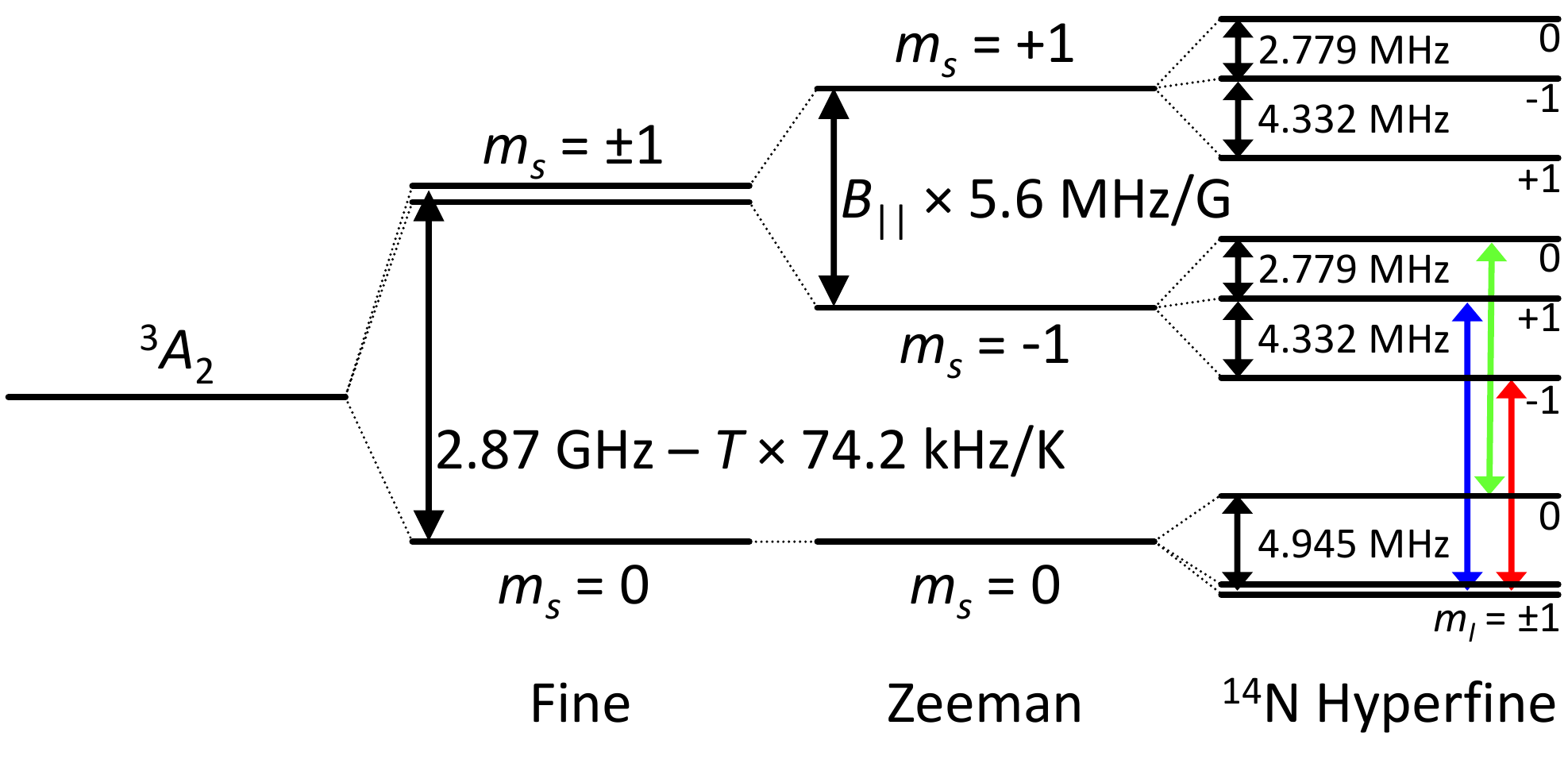}
\put(8,105){\textsf{a}}
\end{overpic}
\begin{overpic}[width=0.48\textwidth]{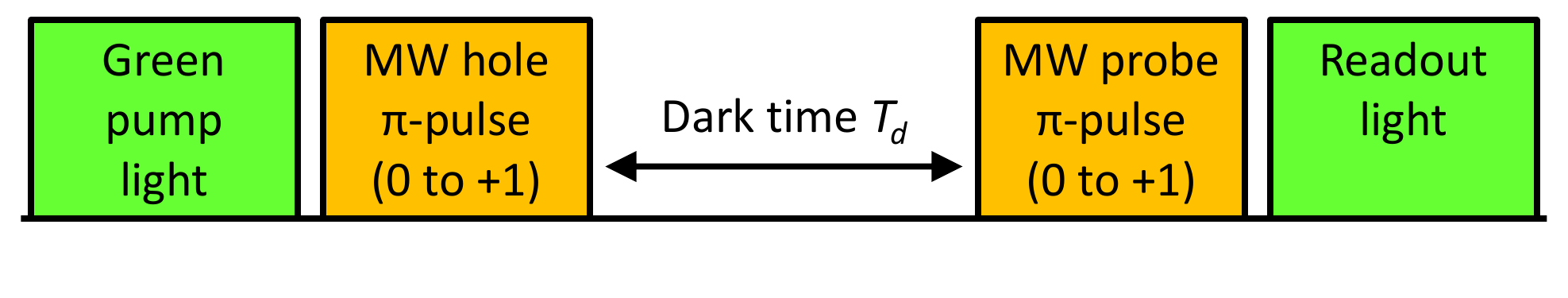}
\put(8,48){\textsf{b}}
\end{overpic}
\begin{overpic}[width=0.48\textwidth]{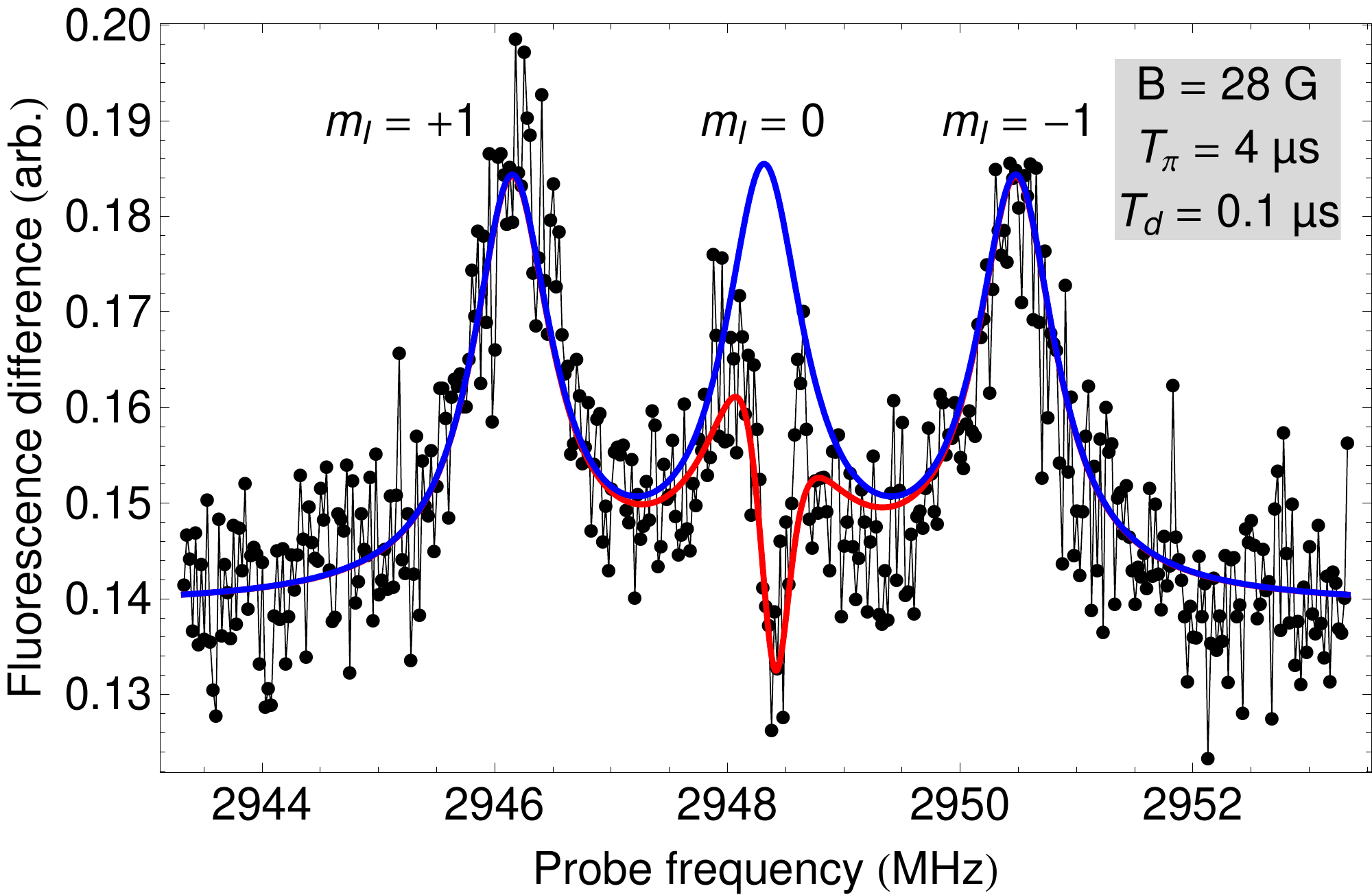}
\put(38,140){\textsf{c}}
\end{overpic}
\end{center}
\caption{\label{fig1}
(a) The \nv\ ground-state energy level structure \cite{steiner_hf, victor_Tdepend}. The $^{14}$N hyperfine interaction is included, but the $^{14}$N Zeeman splitting (0.3 kHz/G) is not. The MW transitions conserve nuclear $m_I$ (colored arrows, also shown in Fig.~\ref{figCWODMR}b).
(b) The pulse sequence used for pulsed hole burning.
(c) The result of a pulsed-hole experiment in sample CVD1 (see Tab.~\ref{fwhmTable}). The 0.3 MHz FWHM hole is Fourier-limited and is narrower than what was achieved with CW hole-burning (Tab.~\ref{fwhmTable}).
}
\end{figure}

The nitrogen-vacancy (NV) color center in diamond is a defect center consisting of a substitutional nitrogen atom adjacent to a missing carbon atom. When negatively charged (\nv), its ground state has electronic spin 1 (Fig.~\ref{fig1}a), and physical parameters such as magnetic field, electric field, and temperature affect the energies of its magnetic sublevels \cite{dima_magnetometry_review, eFieldSensing, victor_Tdepend}. One can measure these parameters by employing optically-detected magnetic resonance (ODMR) techniques \cite{optMagBook, marcus_review}, which use microwave (MW) fields resonant with the NV transitions and detect changes in fluorescence in the presence of excitation light.

The \nv\ ground-state sublevels can be optically accessed and have long spin-relaxation times at room temperature \cite{Nir_CPMG}, making them useful for sensing. When limited by spin-projection noise, the sensitivity is proportional to $\sqrt{\Gamma / N }$, where $\Gamma$ is the ODMR linewidth and $N$ is the number of NV centers probed \cite{dima_magnetometry_review, victor_mag, dreau_powBroad}. In practice, the transitions are inhomogeneously broadened due to differences in the NV local environments, limiting the ensemble sensitivity. Diamond samples with more paramagnetic impurities also have more inhomogeneous broadening, meaning that larger $N$ often comes with larger $\Gamma$. Furthermore, NVs with different Larmor frequencies dephase, which is a limitation in some applications. Although refocusing pulse sequences (such as Hahn echo) can restore the coherence, identifying the sources of ODMR linewidth broadening is essential for NV applications and for understanding the underlying diamond spin-bath and crystal-strain physics. 

In this work we demonstrate novel use of saturation spectroscopy (or ``hole-burning") techniques in an NV ensemble. This is motivated by saturation spectroscopy in atoms, where a spectrally-narrow pump laser selects atoms of a particular velocity class by removing them from their initial state, allowing one to recover narrow absorption lines with a probe laser \cite{rmpHoleBurn}.

We present two hole-burning schemes. The analytically simpler scheme (``pulsed hole-burning") is depicted in Fig.~\ref{fig1}b. This scheme addresses a two-level subsystem ($m_s = 0$ and $+1$) and uses a modified pulsed-ODMR sequence (similar to that of Ref.~\cite{dreau_powBroad}). A spectrally-narrow ``hole" $\pi$-pulse first shelves some NVs into the $m_s=+1$ state, after which a probe $\pi$-pulse reads out its effect on the NV population distribution. Figure \ref{fig1}c shows that using this method can yield hole widths significantly narrower than the inhomogeneous linewidth of this transition.

\begin{figure}[ht]
\begin{center}
\begin{overpic}[width=0.48\textwidth]{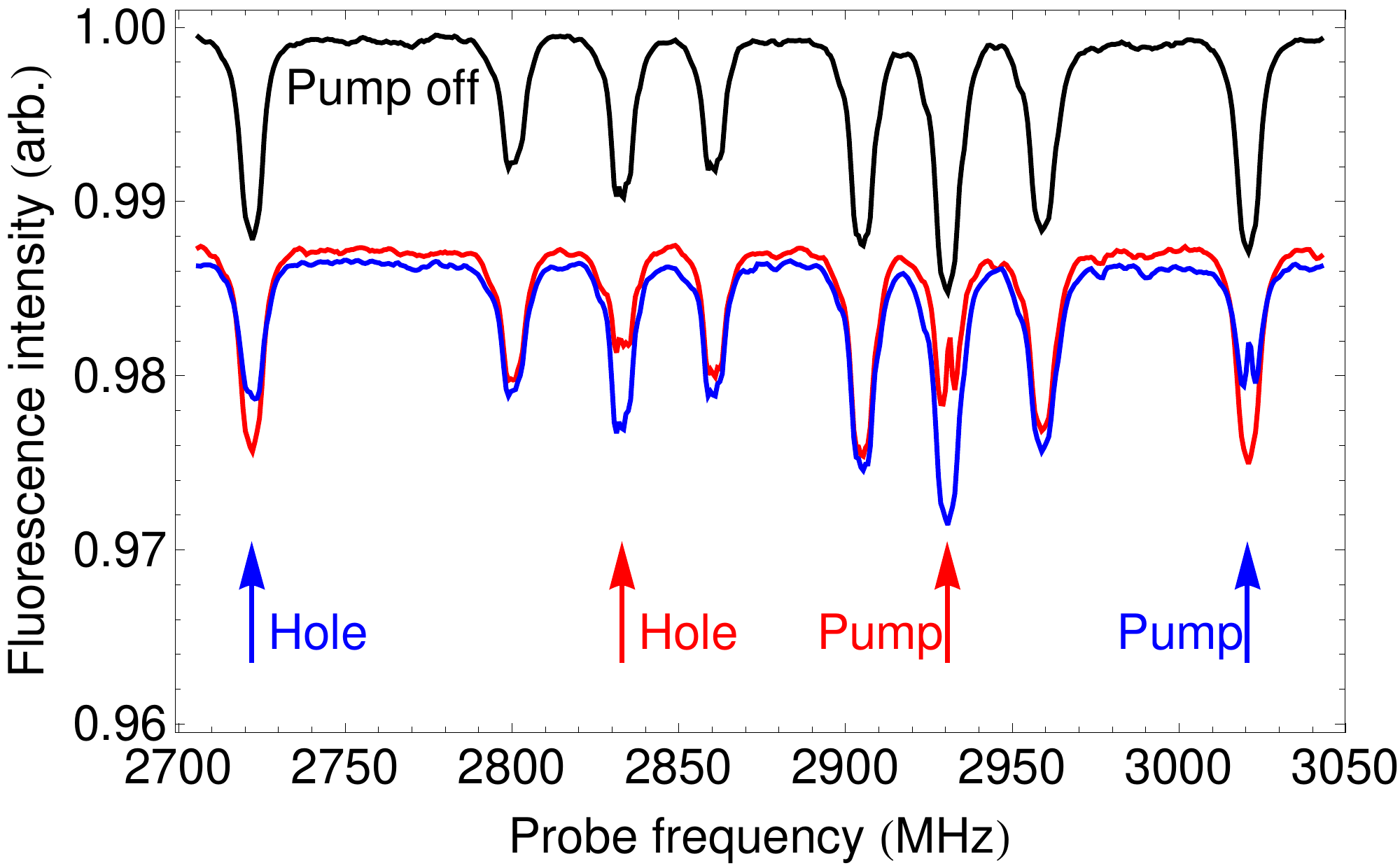}
\put(8,155){\textsf{a}}
\end{overpic}
\begin{overpic}[width=0.48\textwidth]{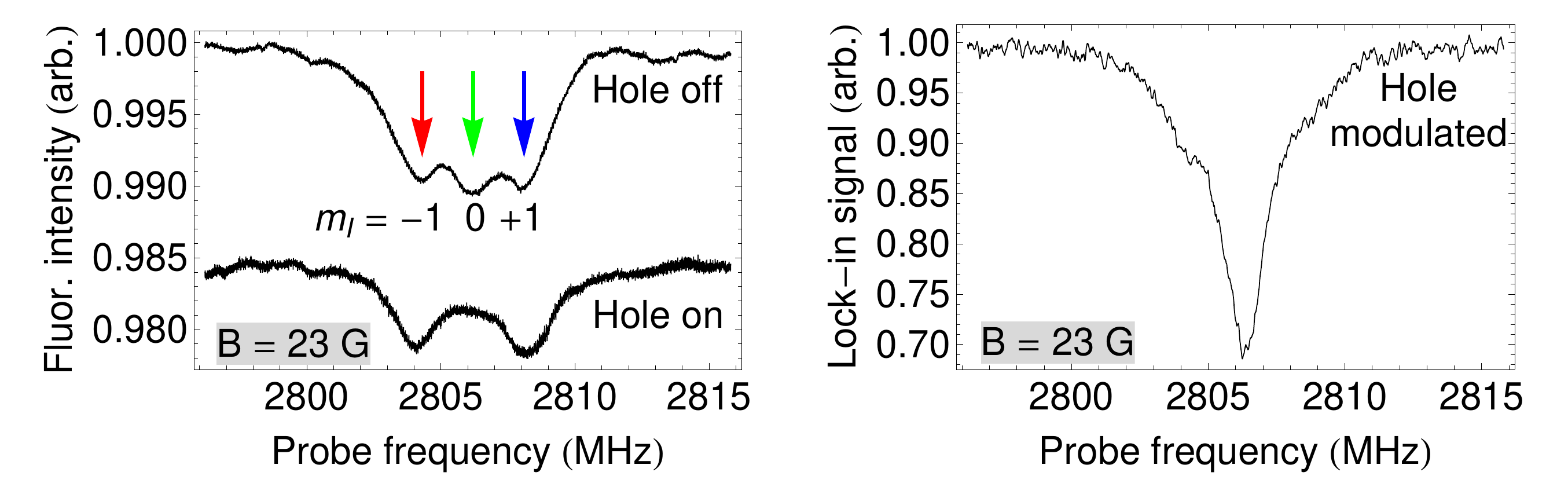}
\put(8,78){\textsf{b}}
\put(233,78){\textsf{c}}
\end{overpic}
\end{center}
\caption{\label{figCWODMR}
(a) CW ODMR spectra in sample HPHT1 with a static magnetic field along an arbitrary direction, which splits the resonances of the four NV-axis alignments into four pairs of frequencies. Burning a hole at any frequency only affects NVs of the associated alignment. The slight asymmetries in ODMR contrast are likely due to differences in MW power delivered to the NVs.
(b) ODMR spectra of the $m_s = 0$ to $+1$ transition, now with an axial magnetic field. This demonstrates the effect of burning a hole in the inhomogeneously broadened transition. The pump frequency is 2934.2 MHz.
(c) Here we amplitude-modulate the pump field in (b) and perform lock-in detection on the NV fluorescence intensity. This reveals the spectrum of NVs affected by the pump field.
}
\end{figure}

The other hole-burning scheme uses all three magnetic sublevels and continuous-wave MW fields (``CW hole-burning"). Here we pump the $m_s = 0$ to $+1$ transition and probe the $m_s = 0$ to $-1$ transition. Again, one observes narrower linewidths than with ordinary ODMR. Figure \ref{figCWODMR} shows ODMR spectra obtained without a CW pump field, with a pump, and a spectrum obtained by modulating the pump amplitude and using lock-in detection. The ODMR linewidth, which is largely determined by inhomogeneous magnetic fields from $^{13}$C nuclei (1.1\% natural abundance) and other sources such as substitutional nitrogen atoms (P1 centers), is reduced in CW hole-burning experiments to a smaller linewidth where the $^{13}$C contribution is removed. We use CW hole-burning to study the causes of ODMR broadening in an NV ensemble and to demonstrate a magnetic-field-insensitive thermometer.

In pulsed hole-burning experiments, we pump and probe the same transition because it yields better fluorescence contrast than with different transitions \cite{suppl}.
This choice is favorable in an NV experiment with poor signal-to-noise. However, applying CW MW fields on the same transition can result in coherent population oscillation (CPO), where the state populations of a quantum system oscillate at the beat frequency between the pump and probe fields \cite{hillmanCPO, WG_cpo}. We use two transitions in CW hole-burning to avoid CPO, and because this choice is useful for determining the dominant source of inhomogeneous broadening. We also show that this more complex hole-burning scheme has applications, a specific example being improved NV thermometry.

Past experiments have demonstrated hole-burning and electromagnetically-induced transparency (EIT) with NV centers (see Refs.~\cite{manson_holeburning, victor_EIT} and references therein) at low temperature using optical fields. Our work is at room temperature, where NVs are most often used in applications, and employs MW fields instead. Moreover, the hole-burning method in Ref.~\cite{manson_holeburning} relies on Raman heterodyne detection and is only feasible at magnetic fields near 1000 G at 5 K temperature. Our schemes are complementary and work at any magnetic field.

\begin{table}[h]
\begin{tabular}{c|c|c|c|c}
Sample	& [N] & [\nv ] 	& ODMR 		& Hole \\
	 	&(ppm)&(ppm)	& FWHM 		& FWHM\\
\hline
HPHT1	& \textless 200& 1-10 & 1.7 MHz & 1.4 MHz\\ \hline  	
HPHT2	& 50 & 1-10 & 1.2 MHz & 0.7 MHz\\ \hline				
CVD1	& 1 & 0.01 & 1.2 MHz& 0.5 MHz CW,\\						
		&   &      &    	& 0.3 MHz pulsed\\	\hline			
CVD2	& \textless 1 & 0.01-0.1 & 0.9 MHz& 0.6 MHz\\ 			
\end{tabular}
\caption{\label{fwhmTable} Details of the diamond samples tested and the smallest linewidths measured (extrapolated to zero MW power). The above widths are for the $^{14}$N hyperfine components of the NV transitions (0.1 MHz accuracy). Each sample has 1.1\% $^{13}$C concentration. HPHT samples were grown with high-pressure high-temperature crystal formation, and CVD samples by chemical vapor deposition.}
\end{table}

\begin{figure}[ht]
\begin{center}
\includegraphics[width=0.48\textwidth]{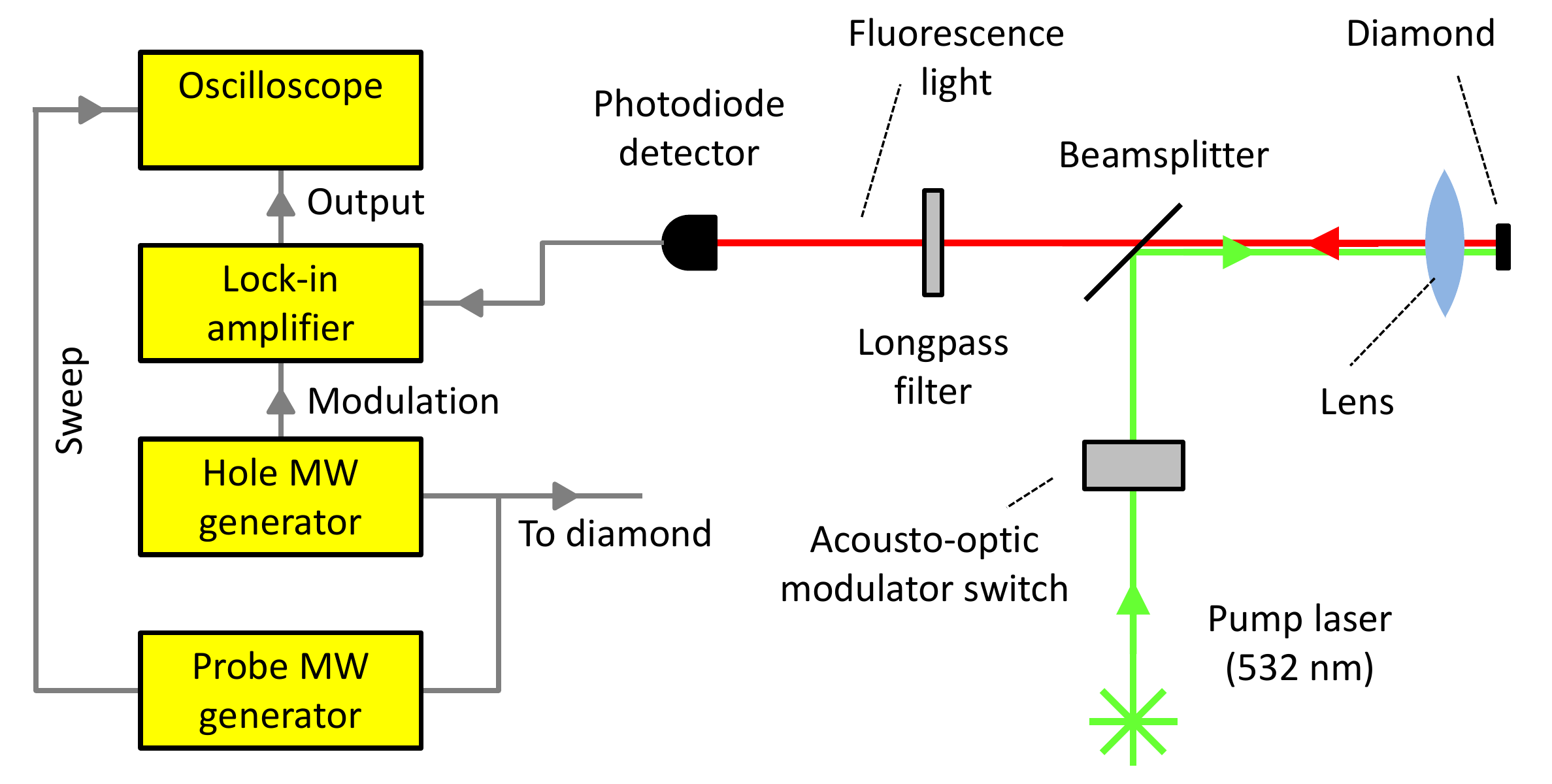}
\end{center}
\caption{\label{figSchematic}
The experimental apparatus for CW hole-burning. We amplitude-modulated the hole MW field at 1.3 kHz. In a pulsed hole-burning experiment, the hole MW field is not modulated and there is no lock-in amplifier.
}
\end{figure}

Figure \ref{figSchematic} shows a confocal microscopy setup, where the NV fluorescence (637-900 nm) is collected with the same lens as is used for excitation and optical pumping (done with 532 nm laser light). We exposed the diamond samples to the pump and probe MW fields with a nearby wire. The pump burns a hole in the $m_s = 0$ population by driving resonant NV centers into the $m_s = +1$ state, which spoils the ODMR contrast. In CW hole-burning measurements, we amplitude-modulated the pump and used lock-in detection to determine the hole linewidths more easily (Fig.~\ref{figCWODMR}c) \cite{foot}. 

\begin{figure}[ht]
\begin{center}
\begin{overpic}[width=0.48\textwidth]{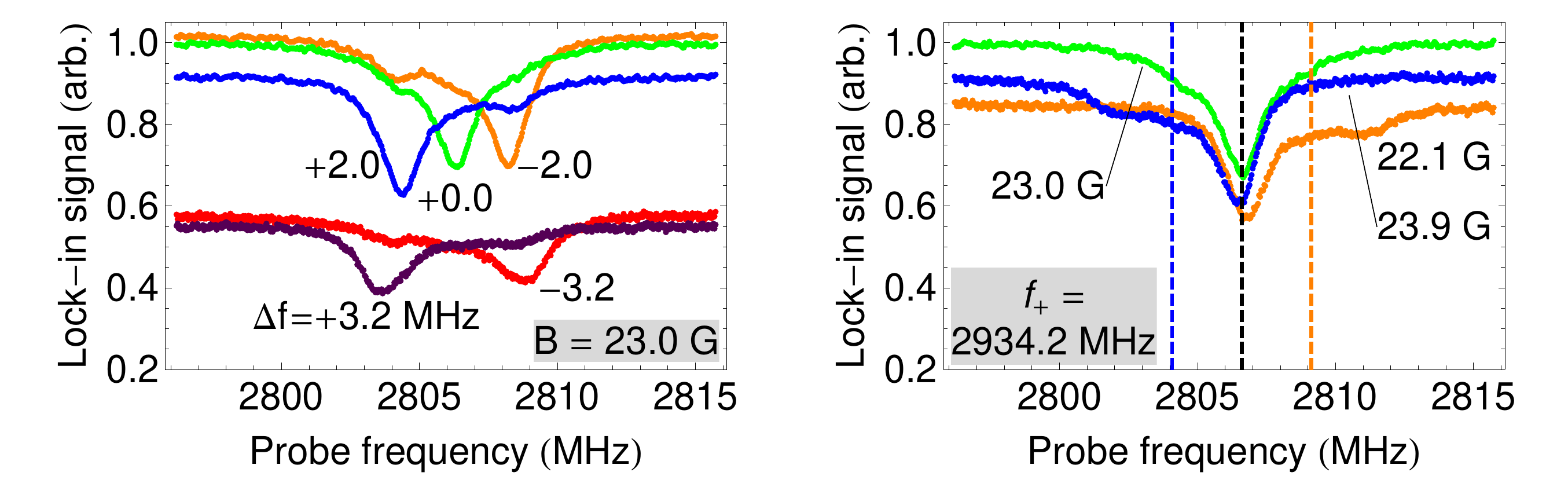}
\put(8,78){\textsf{a}}
\put(131,78){\textsf{b}}
\end{overpic}
\end{center}
\caption{\label{figfB}
(a) The effect of changing the pump frequency ($f_+$ = 2934.2 MHz + $\Delta f$) on the hole center frequency $f_-$ (diamond sample HPHT1). The lock-in signal when the probe is off resonance is due to the NV fluorescence being modulated by the pump MW at the lock-in frequency and is a measure of the pump absorption. Varying $f_+$ confirms the prediction made in Eq.~(\ref{eqB}). (b) Here we keep $f_+$ constant but vary $B$ near 23 G. Frequency pulling causes a $\pm$0.2 MHz shift when $B$ changes by $\pm$0.9 G (dashed lines). Additional data are included in the supplemental material.
}
\end{figure}

The relevant NV electronic ground-state Hamiltonian (in units of hertz) is 
\begin{equation}
H = (D + d^\parallel \delta \epsilon )S_z^2 + \gamma (B+\delta B) S_z,
\end{equation}
where $S_z$ is the dimensionless spin projection operator, $D$ is the zero-field splitting, $\gamma$ = 2.8 MHz/G is the gyromagnetic ratio, $B$ is the axial magnetic field, and $d^\parallel$ is related to the axial electric dipole moment. Each NV has different local axial magnetic field ($\delta B$) and strain ($\delta \epsilon$) values, the distributions of which (with respective widths $\Delta B$ and $\Delta \epsilon$) cause inhomogeneous broadening. A hole-burning test can determine the dominant source of inhomogeneous broadening in a diamond sample. The MW transition frequencies for a specific NV are $f_\pm = D + d^\parallel \delta \epsilon \pm \gamma (B+\delta B)$. If the NV ensemble experiences a distribution of magnetic fields (either from a gradient in the applied magnetic field or from magnetic spins in the diamond) and $\delta \epsilon \approx 0$, then the pump at $f_+$ selects the NVs with a particular $\delta B$ and the spectral hole appears at 
\begin{equation}f_- = 2D - f_+.\label{eqB}\end{equation}
Alternatively, if the ensemble experiences a distribution of axial strains and $\delta B \approx 0$, then $f_+$ selects the NVs with a particular $\delta \epsilon$ and 
\begin{equation}f_- = f_+ - 2 \gamma B.\label{eqD}\end{equation}
Since Eqs.~(\ref{eqB}) and (\ref{eqD}) predict how $f_+$ and $f_-$ are correlated in different physical situations, we can test this correlation to learn whether the effect of $\Delta B$ or $\Delta \epsilon$ is dominant. As seen in Fig.~\ref{figfB}a, varying $f_+$ causes $f_-$ to shift with the opposite sign, meaning that differences in magnetic field cause the inhomogeneous broadening.

Since $f_-$ depends only on $f_+$ and $D$, it is resistant to changes in $B$, as shown in Fig.~\ref{figfB}b. Varying $B$ by $\pm$0.9 G preserves $f_-$ to within $\pm$0.2 MHz while the transition frequencies vary by $\pm$2.5 MHz (shown by dashed lines). If $f_+$ lies on a slope of the $m_s = 0$ to $+1$ lineshape, $f_-$ will be shifted by a ``frequency pulling" effect due to a product of the pump and $m_s=0$ to $+1$ lineshapes. The residual $\pm$0.2 MHz spread in $f_-$ comes from the magnetic field dependence of the $m_s = +1$ sublevel and the frequency pulling effect.

\begin{figure}[ht]
\begin{center}
\includegraphics[width=0.48\textwidth]{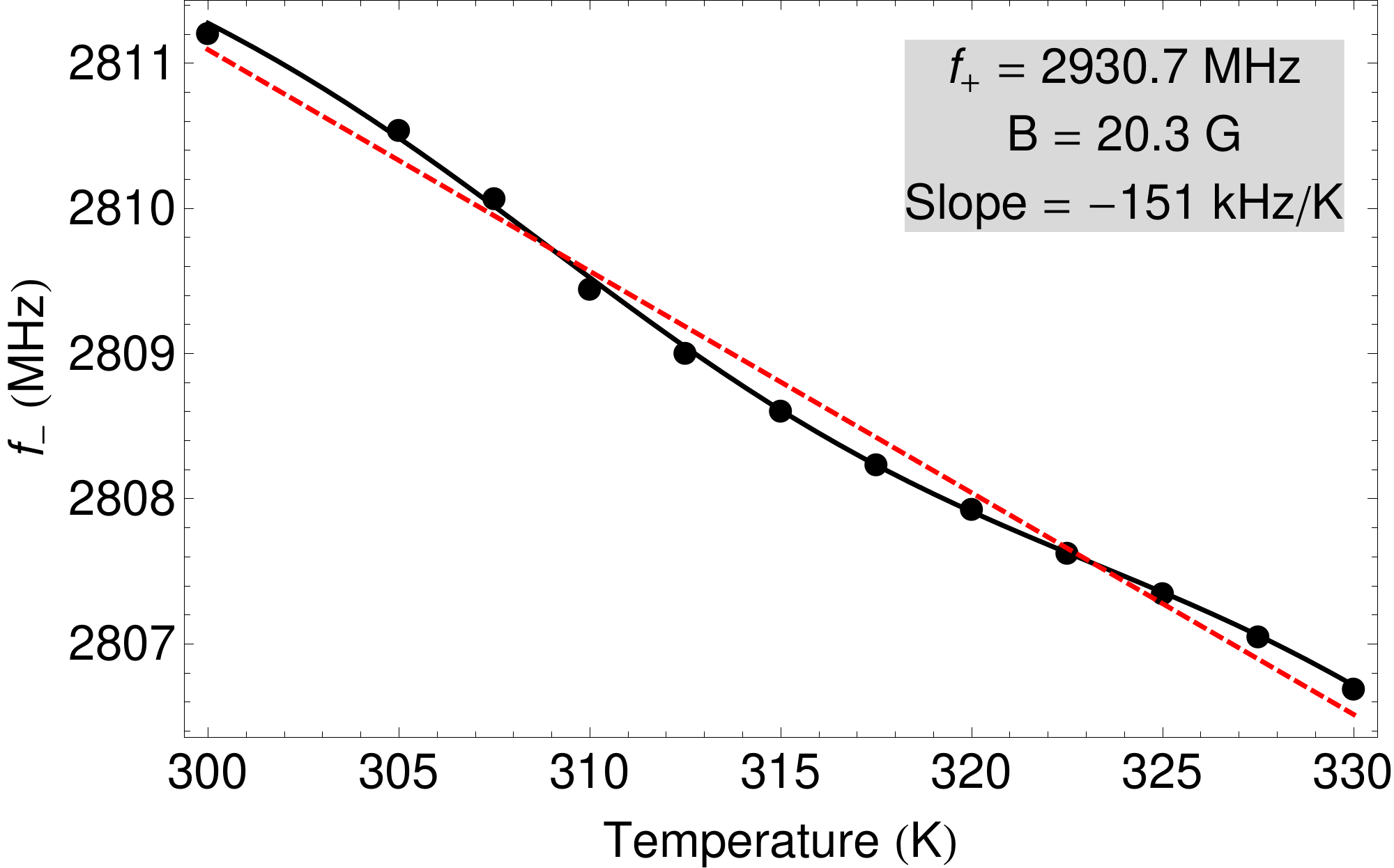}
\end{center}
\caption{\label{figTplot}
Temperature dependence of $f_-$ in sample HPHT2. The solid line is a fit done with Eq.~(\ref{eqFit}), and the dashed line is the linear part of this fit. Note that the oscillation period is not an independent free parameter, but is fixed by the model.
}
\end{figure}

Hole-burning is useful for thermometry because the hole width is narrower than the ordinary ODMR width, $f_-$ is protected against changes in magnetic field, and $f_-$ shifts by twice as much when the temperature changes compared to ordinary ODMR. At room temperature, $D$ shifts by $\alpha = -74.2$ kHz/K \cite{victor_Tdepend}. Using fixed $f_+$ and $B$, we varied the temperature of diamond sample HPHT2 (Fig.~\ref{figTplot}). From Eq.~(\ref{eqB}) we nominally expect $df_- / dT = 2 \alpha$. To anticipate the aforementioned frequency pulling on $f_-$, we used the fit function
\begin{equation}
f_-(T) = c_1 + 2 \alpha T + c_2 \sin{\left[2 \pi \times \frac{\alpha}{A} T + c_3\right]},
\label{eqFit}
\end{equation}
where the $c_i$ are free parameters, $T$ is temperature, and the sine function models the frequency pulling caused by the $^{14}$N hyperfine peaks in the ODMR spectrum (separated by $A$ = 2.166 MHz \cite{steiner_hf}). Repeating this experiment with many $f_+$ frequencies yielded a mean $2 \alpha = -151$ kHz/K with a spread of 2 kHz/K (probably due to the remaining effect of frequency pulling), which is consistent with the expected value of $-148$ kHz/K.

Frequency pulling may limit the thermometer accuracy; however, choosing $f_+$ and $f_-$ wisely can minimize this effect and even improve the sensitivity, as frequency pulling boosts the local slope by up to 30\%. Other methods for making NV magnetometers, thermometers, and clocks more stable in fluctuating thermal and magnetic environments have been developed \cite{fang_Qbeats, ToyliTemperature, cell_thermometry, neumann_nanoT, timekeeping}. These methods use MW pulses to create $m_s = \pm1$ superpositions, canceling deleterious phase accumulation from unwanted temperature or magnetic field drifts. While a CW hole-burning thermometer also uses the $m_s = \pm1$ states, it does not require superpositions or MW pulses.

Figure \ref{figfB}a shows that differences in local magnetic field are the main broadening source of the NV ODMR transitions. The diamond samples contain $^{13}$C nuclei and paramagnetic impurities (such as P1, NV$^0$, and \nv). From their magnetic dipole moments and densities, one can estimate that $^{13}$C nuclei and P1 centers are the main contributors to local magnetic-field inhomogeneity, in roughly equal proportions. These CW hole-burning experiments (with 1.3 kHz modulation frequency) remove the linewidth contribution from $^{13}$C nuclei.

To reach this conclusion, we compared the ODMR and CW hole linewidths in different diamond samples. The $^{13}$C spin bath fluctuates with a correlation time $\tau_c$ of $\sim$10 ms \cite{merilesC13}, which is slow compared to the (1.3 kHz)$^{-1}$ time scale of a CW hole-burning experiment. This means that the $^{13}$C magnetic fields are static for the duration of the experiment. Furthermore, Ref.~\cite{vary_c13_bath} reports that $^{13}$C nuclei are the primary source of broadening ($\sim$0.2 MHz) in samples with few paramagnetic impurities. In Tab.~\ref{fwhmTable}, the hole width is roughly 0.5 MHz smaller than the ODMR width, an improvement which is comparable to what Ref.~\cite{vary_c13_bath} suggests. The 1.3 kHz modulation frequency is slow compared to the P1 $\tau_c$ ($\sim$10 $\upmu$s \cite{deLange2010, deLange2012, nir_spinbath}), meaning P1 centers still contribute to the CW hole linewidth. Since hole burning removes the influence of $^{13}$C nuclei, it effectively leaves a diamond without $^{13}$C. Some experiments avoid the effect of $^{13}$C nuclei with a more ``brute force" approach by using isotopically pure ($^{13}$C-depleted) synthetic diamonds \cite{deltaDoping, nirT2_cold, maurerC13, Mamin_PMMA}.

For short times (faster than any $\tau_c$ and the NV $T_1$), pulsed hole-burning experiments can create spectral holes that are narrower than what was achieved in the slower CW hole-burning experiments described above. Figure \ref{fig1}c shows a 0.3 MHz hole in the ODMR spectrum of sample CVD1, which is narrower than the 0.5 MHz hole we achieved using lock-in detection. This hole width is Fourier-limited and can be reduced to about 0.15 MHz (see supplemental material). If the inhomogeneous magnetic fields are constant for $\tau_c$, the spectral hole vanishes for dark times $T_d > \tau_c$. Extending the $\pi$-pulse duration can reveal what limits the hole width and determine $\tau_c$ for different spin-bath species. Previous experiments have measured NV decoherence in pulsed-microwave experiments to study the P1 and $^{13}$C $\tau_c$ \cite{deLange2010, deLange2012, nir_spinbath, merilesC13}. In comparison, investigating $\tau_c$ with pulsed hole-burning does not require coherent superpositions, which is useful when $\tau_c$ is longer than the NV $T_2$.

In summary, we have demonstrated CW and pulsed hole-burning in NV ensembles in diamond. Using CW hole-burning tests, we distinguished between dominant sources of ODMR broadening and showed that broadening comes mainly from differences in magnetic fields (rather than differences in axial strain). The lock-in detection method eliminates the linewidth contribution from slowly fluctuating $^{13}$C nuclei, while the rapidly-fluctuating magnetic fields from P1 centers and other sources contribute to a reduced linewidth. We also demonstrated a promising temperature sensor that is resistant to magnetic field fluctuations. With pulsed hole-burning, we created narrower spectral holes with Fourier-limited widths, which may be used to study spin-bath dynamics.

NV researchers prefer $^{13}$C-depleted diamond samples, which have better coherence time and sensitivity at the expense of limited availability and greater cost. Since hole burning eliminates the linewidth contribution from $^{13}$C nuclei, this may alleviate the need for $^{13}$C-depleted samples in certain applications. Conversely, hole burning may yield a larger relative improvement in linewidth with $^{13}$C-enriched samples. This may enable high-resolution microwave spectroscopy despite significantly inhomogeneously broadened linewidths, with benefits to $^{13}$C-based NMR and gyroscopy \cite{ran_bulkC13, belth_HH}. In future work, we will search for interactions between NVs with different orientations by pumping NVs in one alignment and probing another, which may help explain the enhanced NV $T_1$ relaxation rate at low magnetic fields \cite{aj_T1}. 

This work was supported by the AFOSR/DARPA QuASAR program, NSF, NATO SFP, and MNSW grant 7150/E-338/M/2013. R.~F.~acknowledges support from the Miller Institute for Basic Research in Science. This work was conducted as part of the Joint Krak\'{o}w-Berkeley Atomic Physics and Photonics Laboratory.

\bibliography{bib_holeBurn}

\begin{thebibliography}{32}
\bibitem [{foo()}]{foot}%
  \BibitemOpen
  \href@noop {} {}\bibinfo {note} {Several lineshape functions (including
  Lorentzian and Gaussian) described our hole spectra well, and we chose to use
  Lorentzian fits. A rigorous analysis of hole parameters (width, contrast, and
  central frequency) should use lineshape functions derived from the broadening
  mechanism.}\BibitemShut {Stop}%
\end{thebibliography}
\onecolumngrid
\newpage


\section*{Supplementary material}

# Supplemental material to "Microwave saturation spectroscopy of nitrogen-vacancy ensembles in diamond"

P. Kehayias, M. Mrózek, V.M. Acosta, A. Jarmola,
D.S. Rudnicki, R. Folman, W. Gawlik, and D. Budker

(Dated: March 9, 2014)

In this supplement we describe and model the NV fluorescence intensity expected in pulsed and CW hole-burning experiments. We provide a model that predicts the NV final-state population fractions, which we use to fit the measurements in Fig. 1c and to estimate the smallest achievable hole width. We also include additional data as an extension to Fig. 4 in the main text.

## 1. FLUORESCENCE IN PULSED HOLE-BURNING EXPERIMENTS - AN INTUITIVE PICTURE

Following Fig. 1 in the main text, we compare the outcomes of two pulsed hole-burning experiments. In the first (experiment 1), the pump and probe drive the $m_s = 0$ to $+1$ transition, while the pump instead drives the $m_s = 0$ to $-1$ transition in the second (experiment 2). Assuming that a $\pi$-pulse resonant with one of the hyperfine components of an NV transition causes a change $\Delta F$ in fluorescence and that the hole frequency $f_h$ is resonant with the $m_I = 0$ hyperfine component, we consider the following illustrative cases for the readout frequency $f_r$ for each experiment:

I  $f_r \approx f_h$.

II  $f_r \neq f_h$ and is not resonant with other NV transitions.

III  $f_r \neq f_h$, but is resonant with other NV transitions.

The final-state populations for the above experiments and cases are portrayed in Fig. S1. Figure S2a shows the result when the hole pulse is disabled, which is identical to an ordinary pulsed-ODMR experiment. When on resonance with a hyperfine component, the probe drives NVs into the $m_s = +1$ state, causing a change $\Delta F$ in fluorescence. In experiment 1,

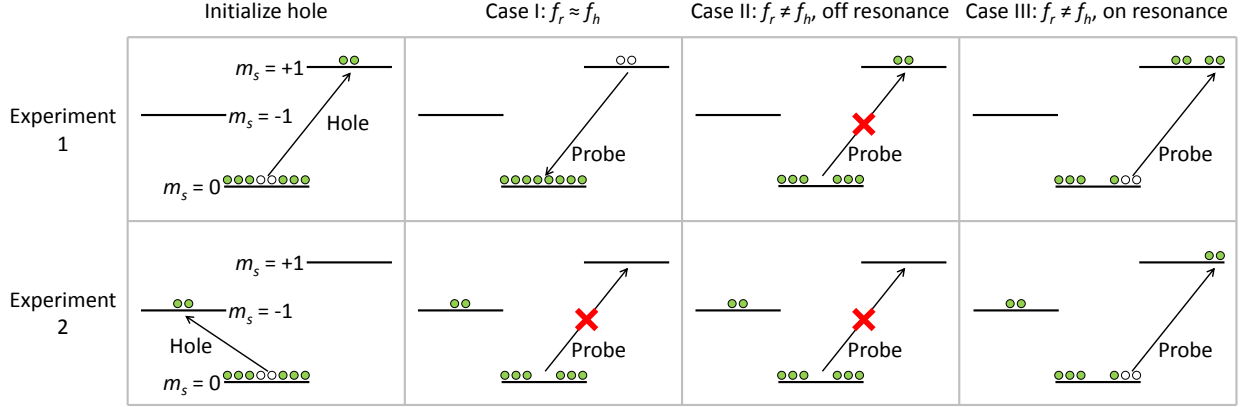

FIG. S1. The actions of the $\pi$-pulses on the sublevel populations. After being initialized into the $m_s = 0$ state, two $\pi$-pulses are applied, with different final-state populations depending on the experiment and $f_r$. Open circles represent NVs removed by the $\pi$-pulse, and filled circles represent NV population after the pulse has acted. A red cross indicates that the probe does not affect the populations.

the probe returns all of the NV population to the $m_s = 0$ state in case I, and the fluorescence is maximal. In case II, the probe has no effect, but the NV fluorescence is reduced by $\Delta F$ since some NVs are in the $m_s = +1$ state. The fluorescence is further reduced in case III, as the probe brings more NVs to the $m_s = +1$ state. Figure S2b shows the expected spectrum.

Experiment 2 has a similar result to that of experiment 1 with one exception. In case I the probe has no effect because the NVs it is resonant with are in the $m_s = -1$ state. Comparing the result (Fig. S2c) to that of experiment 1, we see that experiment 1 is preferable, since the hole pulse has a larger effect on the NV population. We therefore pursued experiment 1 to burn narrow holes, as this scheme has a better signal-to-noise ratio.

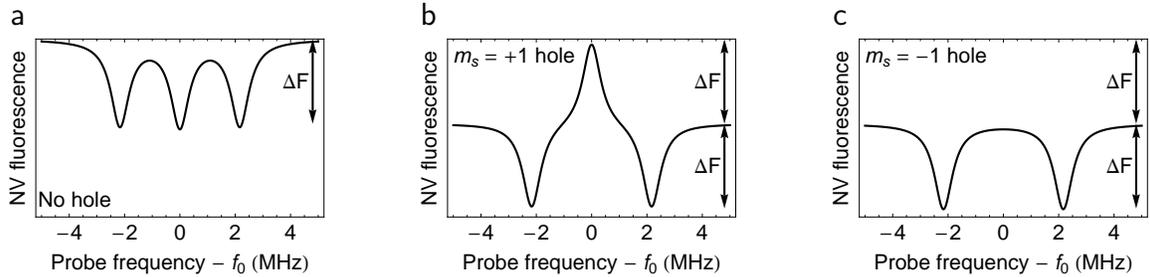

FIG. S2. The expected results of pulsed-ODMR (a) and pulsed hole-burning experiments 1 (b) and 2 (c). We assume that the $m_s = 0$ to $+1$, $m_I = 0$ transition has frequency $f_0$.

## 2. FLUORESCENCE IN CW HOLE-BURNING EXPERIMENTS

The above description also aids in understanding differences in CW hole-burning contrast. Assuming a strong pump (which is chosen to saturate the $m_s = 0$ to $+1$ transition) and a weak probe, we again consider three cases:

I The probe is resonant with either of the MW transitions, and the pump is off.

II The probe is resonant with the $m_s = 0$ to $+1$ transition, and the pump is on.

III The probe is resonant with the $m_s = 0$ to $-1$ transition, and the pump is on.

In case I, the probe suppresses the NV fluorescence by $\Delta F \propto N$, where $N$ the number of resonant NVs in the $m_s = 0$ state. When the pump is switched on, the NV fluorescence is reduced because the pump puts roughly half of the resonant NVs in the $m_s = +1$ state and spoils the optical pumping. The probe has no effect in case II because the $m_s = 0$ to $+1$ transition is saturated. In case III the probe reduces the fluorescence by $\Delta F/2$, since the pump has removed half of the resonant NVs. Although we again see a larger effect when pumping and probing the same transition, analyzing case II is more challenging because of CPO (see references in the main text), which we leave for future work. Instead, we pursued case III as it has interesting physics and applications that are not possible with case II.

## 3. PULSED HOLE-BURNING MODEL

Section 1 provided an intuitive description of the final-state population distributions expected in a pulsed hole-burning experiment. We now form a more complete model to fit pulsed hole-burning data. Furthermore, as the $\pi$-pulse spectral width decreases, the hole width decreases, but since narrow pulses transfer less NV population, the hole also becomes less visible. This model provides an estimate of the smallest achieveable hole width.

For simplicity, we analyze here the two-state pulsed-hole scenario. As the $\pi$-pulse Fourier width is not the main source of limitation, we expect that the results of the simpler scenario are relevant to the more complex CW case as well.

Consider a MW transition ($m_s = 0$ to $+1$, $m_I = 0$) in a pulsed hole-burning experiment. We assume that the sources of inhomogeneous broadening are static. The inhomogeneous

lineshape is a Lorentzian with HWHM $\gamma_i$, and for simplicity we consider the $\pi$-pulse frequency spectrum to also be a Lorentzian with HWHM $\gamma_\pi$. When optically pumped into the $m_s = 0$ sublevel, the probability density function (PDF) for finding an NV with a particular transition frequency $f$ to the $m_s = +1$ state is

$$\frac{1}{\pi}\frac{\gamma_i}{\gamma_i^2 + f^2}. \tag{1}$$

Here we use a frequency scale such that the center frequency of the above Lorentzian corresponds to $f = 0$. A MW $\pi$-pulse, centered at frequency $f_{MW}$, transfers a fraction of the NV population (with resonance frequency $f$) equal to

$$\frac{\gamma_\pi^2}{\gamma_\pi^2 + (f - f_{MW})^2}. \tag{2}$$

Note that the PDF is normalized to have a total population of 1 and the $\pi$-pulse transfers 100% of the NV population that has resonance frequency $f_{MW}$. After the hole $\pi$-pulse (which burns a hole at $f = 0$), the $m_s = 0$ and $+1$ PDFs are

$$p_0^{hole}(f) = \frac{1}{\pi}\frac{\gamma_i}{\gamma_i^2 + f^2} - \frac{1}{\pi}\frac{\gamma_i}{\gamma_i^2 + f^2} \times \frac{\gamma_\pi^2}{\gamma_\pi^2 + f^2}, \tag{3}$$

$$p_{+1}^{hole}(f) = \frac{1}{\pi}\frac{\gamma_i}{\gamma_i^2 + f^2} \times \frac{\gamma_\pi^2}{\gamma_\pi^2 + f^2}. \tag{4}$$

After the readout $\pi$-pulse (with center frequency $f_r$), the PDFs are

$$p_0^{read}(f) = p_0^{hole}(f) + \left[p_{+1}^{hole}(f) - p_0^{hole}(f)\right] \times \frac{\gamma_\pi^2}{\gamma_\pi^2 + (f - f_r)^2}, \tag{5}$$

$$p_{+1}^{read}(f) = p_{+1}^{hole}(f) + \left[p_0^{hole}(f) - p_{+1}^{hole}(f)\right] \times \frac{\gamma_\pi^2}{\gamma_\pi^2 + (f - f_r)^2}. \tag{6}$$

In practice, we vary $f_r$ and determine the population fraction in $m_s = 0$ from the fluorescence intensity. To determine the fraction of NVs in each sublevel as a function of $f_r$ after the readout pulse, we integrate over $f$ to get

$$P_0(f_r) = \frac{f_r^4 \gamma_i + f_r^2\left(\gamma_i^3 + \gamma_\pi \gamma_i^2 + 3\gamma_\pi^2 \gamma_i + \gamma_\pi^3\right) + 4\gamma_\pi^2\left(\gamma_i + \gamma_\pi\right)\left(\gamma_i^2 + \gamma_\pi \gamma_i + \gamma_\pi^2\right)}{\left(\gamma_i + \gamma_\pi\right)\left(f_r^2 + 4\gamma_\pi^2\right)\left(f_r^2 + \left(\gamma_i + \gamma_\pi\right)^2\right)}, \tag{7}$$

$$P_{+1}(f_r) = \frac{\gamma_\pi\left(2f_r^2\left(\gamma_i^2 + 2\gamma_\pi \gamma_i + 2\gamma_\pi^2\right) + f_r^4 + 4\gamma_\pi^2 \gamma_i\left(\gamma_i + \gamma_\pi\right)\right)}{\left(\gamma_i + \gamma_\pi\right)\left(f_r^2 + 4\gamma_\pi^2\right)\left(f_r^2 + \left(\gamma_i + \gamma_\pi\right)^2\right)}. \tag{8}$$

We used a generalized version of Eq. 7, where the hole center frequency is a free parameter, to fit the data in Fig. 1c. The width and contrast of the narrow feature are not free parameters because $\pi$-pulse duration determines $\gamma_\pi$ and the model predicts the contrast. As seen in the figure, the fit describes the hole width and lineshape well. We also investigated a second model where the $\pi$-pulse spectrum is a $\text{sinc}^2$ function instead of a Lorentzian, which more accurately describes our $\pi$-pulses. The Lorentzian and $\text{sinc}^2$ results are similar, meaning the simpler Lorentzian model is sufficient.

As seen in Fig. S3, the effect of the hole on the NV population vanishes as $\gamma_\pi$ approaches 0. To estimate the minimum achievable hole width, we determine $\gamma_\pi$ for which the population deviates by 0.1 when $f_r = 0$. We select 0.1 as a figure of merit; in practice the experimental sensitivity to changes in fluorescence regulates the smallest visible hole width. Solving $P_{+1}(0) = \frac{\gamma_\pi \gamma_i}{(\gamma_\pi + \gamma_i)^2} = 0.1$ yields $\gamma_\pi \approx 0.13\gamma_i$. The widths of the narrow peaks in Fig. S3 are roughly $2\gamma_\pi$ (since there are two $\pi$-pulses). The inhomogeneous FWHM in sample CVD1 (Fig. 1c) is about 600 kHz, meaning the smallest hole we can burn with this method has a FWHM of about 150 kHz.

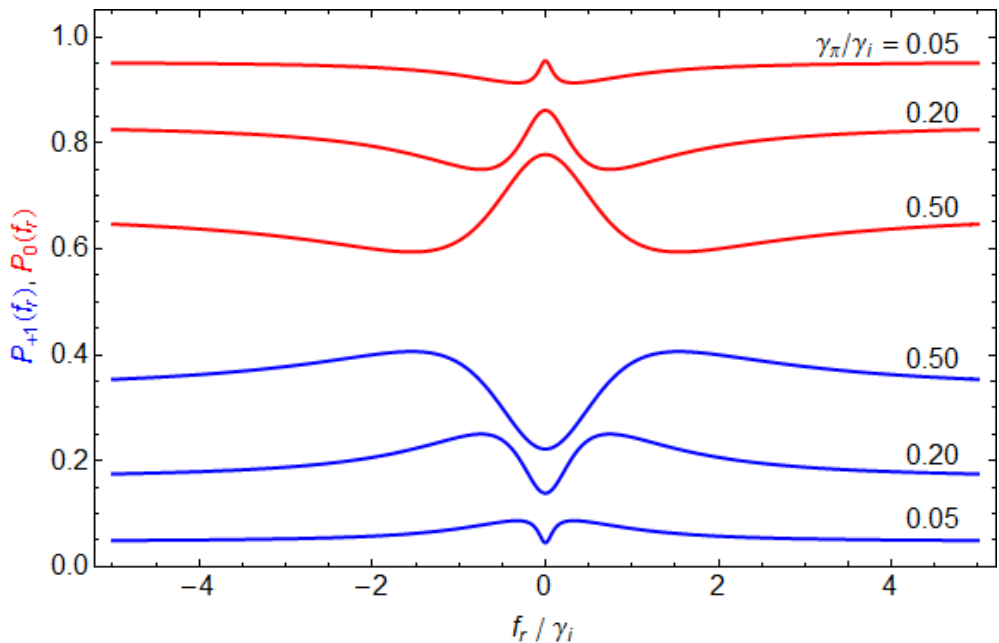

FIG. S3. Final-state population fractions $P_0(f_r)$ and $P_{+1}(f_r)$ for $\gamma_\pi/\gamma_i = 0.05$, 0.20, and 0.5.

## 4. $f_-$ AS A FUNCTION OF $f_+$ AND $B$

As an extension to Fig. 4, we include animated GIF files showing how $f_-$ depends on $f_+$ and $B$. These spectra were taken with sample HPHT2.


\end{document}